\documentclass[sigconf,authorversion]{acmart}
\usepackage{mathpartir}
\usepackage{amsmath,amsfonts}
\usepackage{algorithmic}
\usepackage{graphicx}
\usepackage{textcomp}
\usepackage{xcolor}
\usepackage{hyperref}
\usepackage{multirow}
\usepackage[frozencache]{minted}
\setminted{fontsize=\small}
\usepackage{xspace}

\usepackage{pifont}%
\newcommand{\xmark}{\ding{55}}%

\def\BibTeX{{\rm B\kern-.05em{\sc i\kern-.025em b}\kern-.08em
    T\kern-.1667em\lower.7ex\hbox{E}\kern-.125emX}}
\begin{document}

\newcommand{\PBT}{\textsc{PBT}\xspace}
\newcommand{\PBTs}{\textsc{PBT}s\xspace}

\title{Towards Property-Based Tests in Natural Language
}
\author{Colin S. Gordon}
\orcid{0000-0002-9012-4490}
\affiliation{%
  \institution{Drexel University}
  \city{Philadelphia}
  \state{Pennsylvania}
  \country{USA}
}
\email{csgordon@drexel.edu}

\copyrightyear{2022} 
\acmYear{2022} 
\setcopyright{acmlicensed}\acmConference[ICSE-NIER'22]{New Ideas and Emerging Results }{May 21--29, 2022}{Pittsburgh, PA, USA}
\acmBooktitle{New Ideas and Emerging Results (ICSE-NIER'22), May 21--29, 2022, Pittsburgh, PA, USA}
\acmPrice{15.00}
\acmDOI{10.1145/3510455.3512781}
\acmISBN{978-1-4503-9224-2/22/05}

\begin{abstract}
    We consider a new approach to generate tests from natural language. Rather than relying on machine learning or templated extraction from structured comments, we propose to apply classic ideas from linguistics to translate natural-language sentences into executable tests. This paper explores the application of \emph{combinatory categorial grammars} (\textsc{CCG}s) to generating property-based tests. Our prototype is able to generate tests from English descriptions for each example in a textbook chapter on property-based testing.
\end{abstract}

\maketitle

\section{Introduction}
The ability to specify functional and unit tests using natural language has applications to documenting test cases, tracing requirements to tests, reducing discrepancies between prose functionality descriptions and the properties encoded in test suites, and helping less technical clients build confidence that software is implemented consistently with their understanding of requirements.  It is common to see test suites written using domain-specific languages (DSLs) modeled to look like fragments of English~\cite{mockito}; testing frameworks in the ``Given-When-Then'' behavior-driven development style~\cite{lawrence2019behavior} using specification languages like Gherkin~\cite{wynne2017cucumber} which allow developers to map developer-chosen fixed natural language to sections of test cases; RSpec-style~\cite{chelimsky2010rspec} libraries that have natural language description be embedded in tests; or ad hoc practices in languages like F\#, whose support for (quoted) identifiers containing strings is used in practice to name test cases using natural language descriptions. Prior research has explored variations on these ideas, which can be very effective (some are widely used). But many associate code and descriptions only by convention (F\#, RSpec) or manual processes~\cite{santiago2012generating}, or are limited by a host language (DSLs). Others rely on brittle heuristics to connect language and code, such as: Gherkin; recognizing hard-coded phrases in specific Javadoc clauses~\cite{tan2012tcomment,goffi2016automatic,blasi2018translating}; assuming a restrictive subject-verb-direct-object sentence structure that breaks on many natural sentences~\cite{kamalakar2013automatically}; or using brittle keyword-based heuristics to extract tests for a single testing template~\cite{ansari2017constructing}.
\citet{ahsan2017comprehensive} survey similar techniques.
\citet{sharma2014natural} propose generating tests from a restricted logical representation of requirements that reads similarly to a fragment of English.
While these approaches have notable successes (\citet{motwani2019automatically} generate tests from the official JavaScript specification, exploiting its unusually consistent wording to extract test elements via regular expressions), these are all limited by the lack of a true linguistic model, causing them to fail on novel sentence structures (natural languages are non-context-free~\cite{joshi:91}) and making some extensions non-compositional.

While not yet applied to test generation from text, machine-learning (especially deep learning) is increasingly used to associate code and text~\cite{allamanis2018survey}.  These approaches make good use of large corpora of public software source code to learn rich associations between text and code.  However, a common critique of these techniques is that they are non-modular, making them brittle and difficult to repair. 
Adding an individual word to a neural network-based system~\cite{leclair2019neural,leclair2020improved,hu2020deep,li2020deepcommenter} that was absent in training data is not feasible, nor is correcting handling of individual existing words
The only solution is to to retrain the network.

Classic work in linguistics on compositional models of grammar and sentence meaning offers another, principled way to relate code and natural language, in a way that generalizes to unseen inputs and naturally supports modular extension and bug fixes. We show that it is possible to use a classic linguistic semantic technique to generate property-based tests from natural language. Beyond its direct desirable properties, this approach opens the door to drawing on decades of knowledge from the linguistics community to more effectively generate tests from natural language.

\section{Categorial Grammar, Semantic Parsing, and Property-Based Tests}
\begin{figure*}\small
\begin{mathpar}
\inferrule*[right=$<$]{
\inferrule*[right=lex]{ }{\textrm{3}\vdash NP \Rightarrow 3}\\
    \inferrule*[right=$>$]{
    \inferrule*[right=lex]{ }{\textrm{is}\vdash (S\setminus NP)/ADJ\Rightarrow (\lambda p\ldotp\lambda n\ldotp p~n)}\\
    \inferrule*[right=lex]{ }{\textrm{even}\vdash ADJ \Rightarrow (\lambda n\ldotp n\%2=0)}
    }{
        \textrm{is}~\textrm{even}\vdash (S\setminus NP)\Rightarrow(\lambda n\ldotp n\%2=0)
    }
}{\textrm{3}~\textrm{is}~\textrm{even}\vdash S \Rightarrow 3\%2=0}
\end{mathpar}
\vspace{-2em}
\caption{A derivation translating ``3 is even'' to the proposition $3\%2=0$.}
\vspace{-1em}
\label{fig:demo}
\end{figure*}

Linguists have long studied precise models of grammar and meaning, producing rich literature on computing natural language text's semantic meaning from a grammatical parse tree --- the parse guides how the logical meanings of individual words are combined to compute the meaning of a whole sentence~\cite{barker2007direct}.
\emph{Combinatory Categorial Grammars} (\textsc{CCG}s)~\cite{steedman2012taking,Baldridge:2003:MCC:1067807.1067836} are one established body of techniques to model both parsing and translation into a \emph{logical form} representing sentence meaning, a task known as \emph{semantic parsing}.
\textsc{CCG}s are theoretically powerful (mildly-context-sensitive~\cite{VijayEq:94,Kuhlmann2015}) and have been shown to capture general accounts of subtle linguistic phenomena in a number of natural languages, well enough to parse large corpora of English~\cite{hockenmaier2007ccgbank}, German~\cite{hockenmaier2006creating}, Hindi~\cite{ambati2018hindi}, and Japanese~\cite{mineshima2016building}.
\textsc{CCG}s can model meaning using any lambda calculus with a boolean-like type (technically, a Heyting Algebra)~\cite{lambek1988categorial}, so can be used to generate meanings in many logics or programming languages.
\looseness=-1

\emph{Property-based tests} (\PBTs)~\cite{claessen00quickcheck} 
are a form of guided random testing, focusing on properties that should hold for \emph{classes} of inputs rather than individual inputs. Inputs are produced by combining and filtering primitive random input generators, and transforming results of other generators.
\PBTs generally take the form of a universal property \mintinline{javascript}{forall(g,f)}, where \mintinline{javascript}{g} is a generator and \mintinline{javascript}{f} is a function which either returns a boolean or performs assertions to check a property of any input produced by \mintinline{javascript}{g}.
When run, the test draws many random inputs from \mintinline{javascript}{g}, calling \mintinline{javascript}{f} on each, and fails if \mintinline{javascript}{f} ever returns false or fails an assertion.
Typical \PBT implementations make heavy use of anonymous functions, and represent properties with a datatype which carries operations corresponding to conjunction, disjunction, negation, and implication of properties. These datatypes (representing logical claims) are then adequate to serve as a target semantics for \textsc{CCG}-based semantic parsing, meaning \textsc{CCG}s can be used to translate natural language to property-based tests.\footnote{In principle \textsc{CCG}s could be used for regular test assertions as well, but we focus on \textsc{PBT}s because they specify complete tests.} The rest of this section outlines how this can work.

Categorial grammars annotate each word with a set of grammatical \emph{categories} describing how it combines with other words and clauses. The supported categories include both primitive categories, and categories modeling words whose semantics take arguments.  For English the base categories typically include sentences ($S$), noun phrases ($NP$), and adjectives ($ADJ$), among others.  The (language-independent) non-primitive categories are built from left and right \emph{slash types}.  A left slash type $A\setminus B$ is the category of sentence fragments whose composition with a fragment of type $B$ on its left result in a larger fragment of grammatical type $A$. A right slash type $A/B$ expects the $B$ to the right.\footnote{In both cases the result category is on the left, the argument is on the right, and the top of the intervening slash ``leans'' in the direction of the argument.}
These intuitions, and other means of combining grammar sentence fragments, are captured by inference rules specifying how adjacent sentence fragments interact.
For example:
\vspace{-0.75em}
\begin{mathpar}\small
    \inferrule[Right-Application ($>$)]{\Gamma\vdash X/Y \Rightarrow f \\ \Delta\vdash Y \Rightarrow a}{\Gamma,\Delta\vdash X \Rightarrow f~a}\;\;
    \inferrule[Left-Application ($<$)]{\Gamma\vdash Y \Rightarrow a \\ \Delta\vdash X\setminus Y \Rightarrow f}{\Gamma,\Delta\vdash X \Rightarrow f~a}
\end{mathpar}
\vspace{-1.25em}\\
Here $\Gamma$ and $\Delta$ are non-empty sequences of words, $X$ and $Y$ (and the slash types) are grammatical types, and $f$ and $a$ are the semantics (classically, logical denotation) of the individual fragments.
In the first rule, $\Gamma$ is a sentence fragment that is nearly of grammatical type $X$, if it only had a $Y$ to the right, so its logical form is a function $f$. Applying that function to the semantics of $\Delta$ (whose type is the needed $Y$) yields semantics for a grammatical phrase of type $X$.
The second rule is symmetric.  These rules plus assumptions about individual words is enough to build a kind of logical derivation
that parses a simple sentence into its meaning as in Figure \ref{fig:demo}. There, ``is'' combines first with its right argument ``even'' (an adjective ADJ), yielding sensible semantics for verb phrase ($S\setminus NP$) ``\textunderscore\ is even'': $(\lambda n\ldotp n\%2=0)$. Then the result combines on the left with ``3'' (a noun phrase NP), completing the sentence.
Note that it must be possible to formulate false claims (failing tests).

Assumptions about individual words form a \emph{lexicon}: a set of grammatical categories and semantics for each word.  A word with multiple meanings may have multiple lexicon entries.
\textsc{CCG}s isolate knowledge for specific natural languages to this lexicon, reusing the core rules across any natural language.  \emph{Wide-coverage \textsc{CCG} lexicons} capable of correctly parsing large text corpora exist for English~\cite{hockenmaier2007ccgbank}, Hindi~\cite{ambati2018hindi}, German~\cite{hockenmaier2006creating}, and Japanese~\cite{mineshima2016building}.
The natural modular structure of lexicons means that these existing lexicons can be directly extended with domain-specific terminology.
\looseness=-1

For a more complex example than Figure \ref{fig:demo}, consider a \PBT for ``every even integer is divisible by 2.'' This quantifies over all \emph{even} integers from a generator, which is typically expressed in code by applying a filter operation to a generator. 
Our lexicon entry for the quantifier ``every'' captures this:
\looseness=-1
\vspace{-0.5em}
\[\begin{array}{l}
\textrm{every}\vdash ((S/(S\setminus NP))/CN[Gen])/ADJ \Rightarrow\\
\mintinline{javascript}{P=>gen=>claim=>fc.property(gen.filter(P),claim)}%
\end{array}
\vspace{-0.5em}
\]
This defines ``every'' as a word which, given an adjective ($P$), a common noun ($gen$) of a particular flavor (see Section \ref{sec:lingfindings}), and a sentence fragment ($claim$) that is intuitively ``missing its subject'' (and looking for a suitable subject noun phrase to its left), yields a sentence. The semantics of that sentence (given using Javascript anonymous functions and a specific \PBT library) is a property test that asserts that $claim$ holds for all inputs produced by the generator $gen$ corresponding to that common noun, for which the adjective $P$ is accurate (implemented by filtering the generator).

\textsc{CCG}s include, \emph{and our prototype uses}, 6 additional rules to extend coverage to linguistic constructs like long-distance dependencies~\cite{morrill1995discontinuity,moortgat1996generalized}, unlike-coordination~\cite{bayer1996coordination,carpenter1997type}, and cross-serial dependencies~\cite{shieber1985evidence} (which are non-context-free); and extended slash types~\cite{steedman2012taking,Baldridge:2003:MCC:1067807.1067836} that capture how individual words restrict reorderings of other phrases (e.g., island constraints~\cite{fodor1983phrase}).
The established use of \textsc{CCG}s to analyze many subtle linguistic constructions~\cite{steedman2012taking,jacobson1999towards} in many languages~\cite{hockenmaier2007ccgbank,hockenmaier2006creating,ambati2018hindi,mineshima2016building}, and their demonstrated utility in parsing corpora like the \emph{Wall Street Journal}~\cite{hockenmaier2007ccgbank} and \emph{Alice in Wonderland}~\cite{yeung-kartsaklis:2021:SemSpace} strongly suggest that in contrast to popular heuristic approaches to generating tests from text (mentioned in the introduction), \textsc{CCG}s will impose no fundamental restrictions on the language used in test specifications --- all restrictions on language used for tests in our approach would stem from the lexicon used, which as discussed can be modularly extended or improved over time.
\looseness=-1

\setmintedinline{fontsize=\footnotesize}
\begin{table*}[h!]
    \small
\caption{\textsc{STTP} test specifications with variants handled by our prototype.}
\setlength{\tabcolsep}{3pt}
\vspace{-1em}
\begin{center}
\begin{tabular}{|c|l|}
\hline
\# & Original and Modified Test Specification, (Beta-Reduced) Logical Representation  \\ \hline
\multirow{3}{*}{1} & For all floats, ranging from 1 (inclusive) to 5.0 (exclusive), the program should return false \\
& $\leadsto$ Any float greater than or equal to 1 and less than 5 is not passing \\
& $\hookrightarrow$ \verb+foreach(filter(floats,\x.((lessthan(1,x) | equals(1,x)) & lessthan(x,5))),\n.-passing(n))+ \\
\hline
\multirow{3}{*}{2} & For all floats, ranging from 5 (inclusive) to 10 (inclusive), the program should return true \\
& $\leadsto$ Any float greater than or equal to 5 and less than or equal to 10 is passing \\
& $\hookrightarrow$ \verb+foreach(filter(floats,\x.(lessthanoreq(5,x) & lessthanoreq(x,10))),\n.passing(n))+\\
\hline
\multirow{3}{*}{3} & For all invalid grades (which we define as any number below 0.9 or greater than 10.1), the program must throw an exception \\
& $\hookrightarrow$ For any float less than 1 or greater than 10 passing throws an exception\\
& $\hookrightarrow$ \verb+foreach(filter(floats,\x.(lessthan(x,1) | lessthan(10,x))),\x.checkthrows(isexc,\u.passing(x)))+\\
\hline
\multirow{3}{*}{4} & For all numbers divisible by 3, and not divisible by 5, the program returns ``Fizz''  \\
& $\leadsto$ For any number divisible by 3 and not divisible by 5 fizzbuzz returns ``Fizz'' \\
& $\hookrightarrow$ \textcolor{red}{\xmark}\verb+foreach(filter(integers,\x.(divisibleby(x,5) & -divisibleby(x,3))),\x.(fizzbuzz(x)=Fizz))+\\
\hline
\multirow{3}{*}{5} & For all numbers divisible by 5 (and not divisible by 3), the program returns ``Buzz.''  \\
& $\leadsto$ For any number divisible by 5 and not divisible by 3 fizzbuzz returns ``Buzz''  \\
& $\hookrightarrow$ \textcolor{red}{\xmark}\verb+foreach(filter(integers,\x.(divisibleby(x,5) & -divisibleby(x,3))),\x.(fizzbuzz(x)=Buzz))+\\
\hline
\multirow{2}{*}{6} & For all numbers divisible by 3 and 5, the program returns ``FizzBuzz''.$\leadsto$ For any number divisible by 5 and 3 fizzbuzz returns ``FizzBuzz''  \\
& $\hookrightarrow$ \textcolor{red}{\xmark}\verb+foreach(filter(integers,(\x.divisibleby(x,5) & divisibleby(x,3))),(\z.fizzbuzz(z)==FizzBuzz))+\\
\hline
\multirow{2}{*}{7} &  The program throws an exception for all numbers that are zero or smaller.$\leadsto$ For any number less than or equal to zero fizzbuzz throws an exception\\
& $\hookrightarrow$ \verb+foreach(filter(integers,\x.(lessthan(x,0) | equals(0,x))),\x.checkthrows(isexc,\u.fizzbuzz(x)))+\\
\hline
\end{tabular}
\end{center}
\label{tbl:sttp}
\vspace{-1em}
\end{table*}

\section{Prototype Implementation}
We have used NLTK~\cite{bird2006nltk} to implement a publicly-available prototype~\cite{icsenierdata22} that translates English to property-based tests in Javascript using the \texttt{fast-check} \PBT library~\cite{fastcheck}. NLTK includes an implementation of semantic parsing using \textsc{CCG}s: given a set of base categories and lexicon entries, the library can construct a chart parser~\cite{Clark:2002,Clark04theimportance,Clark:2003} for the specified lexicon.
NLTK's \textsc{CCG} support assumes a simply typed lambda calculus for semantics; 
we generate semantics in that form with \texttt{fast-check} identifiers, and rewrite the syntax into JavaScript syntax after the fact.
The prototype is primarily the lexicon itself. We constructed the lexicon by hand-writing entries for words present in our evaluation study based on a combination of examining the samples in our corpus, prior experience describing property-based tests (e.g., in teaching), and background knowledge about the treatment of certain English-language grammatical phenomena in \textsc{CCG}s and related grammar formalisms, such as quantification~\cite{steedman2012taking,jacobson1999towards}, and coordination (``and'' and ``or'')~\cite{carpenter1997type,morrill2012type}.

This initially manual authoring process highlights some strengths relative to the increasingly popular use of deep neural networks for similar tasks~\cite{allamanis2018survey}.  As noted earlier, neural techniques are non-modular and do not permit fixes for individual words. By contrast, lexicons, while intricate, are inherently modular. Several times while working on the evaluation below, we wrote incorrect semantics or incorrect grammatical types for several words.  Fixing them was a simple matter of correcting individual entries, which had no effect on examples that already worked but did not involve the word in question. The modular construction extends to the addition of new words.
Using \textsc{CCG}s, adding basic negation support is a simple matter of adding one lexicon entry: $\textrm{not}\vdash ADJ/ADJ\Rightarrow \mintinline{javascript}|P=>x=>!P(x)|$.  
More complete support for negation in natural language requires additional lexicon entries, but here again \textsc{CCG}s offer advantages over neural networks: detailed \textsc{CCG} treatments of negations in natural language already exist~\cite{steedman2012taking}, while neural networks still struggle to handle language surrounding boolean operations~\cite{traylor2021and,ettinger2020bert}.
\looseness=-1

We expect that in practice, adding support for new properties and data types (e.g., packaging extensions for a specific library or program) will require \textsc{CCG}'s strong support for modularity.
In the future we plan to adapt techniques for \emph{learning} \textsc{CCG} lexicons~\cite{Zettlemoyer:2005,artzi2013weakly} to extend an existing English lexicon~\cite{hockenmaier2007ccgbank}, but those results will retain the ability to individually extend or correct individual words.
\looseness=-1

\section{Experience With The Prototype}
\label{sec:eval}
To evaluate basic feasibility of our approach, we collected complete sentences describing properties paired with corresponding \PBTs, from books teaching \PBT.
To avoid biasing results towards what we thought would be easier to support, we sought cases where an existing text had a pairing of a \PBT with a sentence describing what the test checked. We surveyed 6 books covering property-based testing~\cite{sttp,o2008real,hinojosa2013testing,lundin2015testing,wampler2014programming,chiusano2014functional}, locating 26 such pairs.
We translated 7 descriptions from Aniche's \emph{Software Testing: From Theory to Practice} (\textsc{STTP})~\cite{sttp} into Javascript \texttt{fast-check} tests, with mild rephrasing to replace use of ``the program'' with the specific function being tested and to work around limitations of the NLTK frontend (lack of support for punctuation and numerals).
We then studied the 19 other descriptions looking for signs of additional linguistic complexity beyond what is treated in existing texts on categorial grammars for English~\cite{steedman2012taking,carpenter1997type,morrill2011categorial,jacobson2014compositional}.

\paragraph{Basic Feasibility}
Table \ref{tbl:sttp} shows the original text, rephrasings, and computed $\lambda$-calculus logical forms for the 7 \textsc{STTP} tests, which the prototype converts to Javascript.

\paragraph{Adequacy of Generated Tests}
Each generated test correctly formalizes the corresponding text.
3 of the \textsc{STTP} exemplars (marked with \textcolor{red}{\xmark}) fail when run on a correct implementation because they were under-specified.
Test \#7 requires an exception for negative numbers and 0. But these can also be divisible by 3 or 5, conflicting with tests \#4--6. The textbook's solutions filter generators to produce only inputs are greater than or equal to 1 for those tests. By adding a lexicon entry for ``positive'' we can generate tests equivalent to the solutions (using ``positive numbers''). This kind of detail is both the sort of detail sometimes omitted intentionally in informal prose, but also the sort of detail often omitted \emph{unintentionally} when under-specifying.
Directly connecting the English to the tests revealed the English was imprecise about expected behavior ---  exactly the sort of divergence between natural language specifications and tests we sought to identify.

Natural language often admits multiple parses for the same sentence, which occasionally yields different logical forms. The sentences in Table \ref{tbl:sttp} yielded respectively 20, 4, 6, 45, 45, 3, and 1 distinct parse trees, but all parses of each sentence produced equivalent logical forms (e.g., differing only in the names of bound variables).

\paragraph{Lexicon Size}
Parsing these examples containing 29 words and numbers requires 46 lexicon entries.
We believe this is reasonable.
Only 3 rules are specific to the subjects under test: the entries for ``passing'' and ``fizzbuzz'' (there are two entries for ``passing'', usable as a function or an adjective).
7 rules are working around limitations of NLTK, which does not have generic handling of numerals (4) of string literals (3). A more robust implementation would have generic handling of numeric constants and string literals.
Thus there are 36 reusable lexicon entries for 19 reusable words. We would expect lexicon size to be roughly linear in the number of words handled, and clearly 19 entries is a lower bound for 19 words. Duplicates handle words that may be used in multiple grammatical roles with similar semantics.

\paragraph{Linguistic Flexibility}
Even these 7 tests involve a range of subtle and commonly-occurring linguistic phenomena, such as overloading ``and'' and ``or'' to coordinate multiple arguments (``divisible by 3 and 5'') or multiple adjectives (``divisible by 3 and not divisible by 5'').
Moreover, because the approach is built on \textsc{CCG}s and therefore directly model compositional meaning of natural language, we can have high confidence that additional sentences using the same words in known grammatical roles will also parse.
As an example of this in tandem with modular lexicon extension, adding an entry for ``is'' allows generalizing already to ``any float that is passing is greater than or equal to 5 or less than or equal to 10'' --- a plausible additional description which strengthens test \#2 to imply that only numbers in that range are passing.%

\paragraph{Linguistic Findings}\label{sec:lingfindings}
Most of our grammatical types and semantics follow standard linguistic models.
We did, however, find that using \PBTs as linguistic semantics required one additional grammatical distinction in common nouns, between those used to constrain the domain of a property (and therefore correspond to constraining \emph{generators}) and those which play a role in evaluating the property (and therefore correspond to \emph{predicates}).
For example, in considering ``Every integer is an integer'' as a \PBT description, the English common noun ``integer'' corresponds once to a generator of integers (as the domain of the quantifier ``every'') and once to a predicate on values (which is true only for integers), which would manifest as two lexicon entries for ``integer'':
\vspace{-0.5em}
\[\begin{array}{l}
    \textrm{integer} \vdash CN[Gen] \Rightarrow \mintinline{javascript}{fc.integer()}\\
    \textrm{integer} \vdash CN[Chk] \Rightarrow \mintinline{javascript}{Number.isInteger}
\end{array}
\vspace{-0.5em}
\]
Our prototype lexicon repeats this distinction with other common nouns to distinguish value checks from generators.
This distinction refines standard grammars, similar to a distinction made for mathematical text~\cite{ranta1994syntactic,ranta1995context}, where additional mathematically-required subcategorizations yield more precise grammars.
\looseness=-1

In examining the additional 19 \PBTs from other texts~\cite{o2008real,hinojosa2013testing,lundin2015testing} (available with our prototype~\cite{icsenierdata22}), every sentence structure is explained by standard grammatical theories, and can be parsed by a grammar-only \textsc{CCG} parser~\cite{yoshikawa-etal-2017-ccg}.
We have not implemented semantics for these tests because they require resolution of definite referents, a linguistic construct with subtle semantics~\cite[Ch.\ 7]{carpenter1997type}, \cite[Ch.\ 7]{steedman2012taking} for which we would prefer to import existing solutions~\cite{abzianidze2017parallel}.
These additional examples, however, reveal the limitations of focusing on educational material on \PBT to find strict English-to-test correspondences.
Two of the additional sources specify nearly the same tests for sorting a list~\cite{lundin2015testing,o2008real}, one repeats the same FizzBuzz underspecification of \textsc{STTP}~\cite{lundin2015testing}, and one explicitly states only trivial properties~\cite{hinojosa2013testing} (about adding integers), suggesting work is needed to develop suitable evaluation corpora.

\paragraph{Limitations}
Our prototype's primary limitation is lexicon size; words absent from the lexicon cannot be used. Fortunately the lexicon is inherently modular, and for reasons outlined below (Section \ref{sec:future}) we believe growing this lexicon is feasible.
In addition, our extraction to JavaScript is currently text-based (e.g., textually replacing ``\texttt{forall}'' with ``\mintinline{javascript}{fc.property}'' because NLTK does not allow periods in logical forms). Future improvements should replace this with a proper traversal of the logical form's abstract syntax tree.

\section{Future Work}\label{sec:future}
The key piece of this approach is the lexicon, which clearly must be extended beyond our prototype. Fortunately, there is already a wide-coverage CCG lexicon for English, \textsc{CCGBank}~\cite{hockenmaier2007ccgbank,hockenmaier2005ccgbank}, which is able to parse a 4.5 million-word corpus~\cite{Marcus:1993} extracted from the \emph{Wall Street Journal}. Though this certainly lacks many software-related terms and includes many terms irrelevant to testing, its more than 74,000 entries cover many different uses of common English words, including quantifiers, prepositions (\emph{from}, \emph{of}, \emph{as}, \ldots), and distinctions between various verb classes (e.g., with or without direct objects). While some of these will require testing-specific semantics, their grammatical categories can be reused directly, cutting down the most laborious part of lexicon creation. Smaller cross-linguistic lexicons~\cite{abzianidze2017parallel} offer a starting point to extend this beyond English. These can be supplemented by work to {learn} further lexicon extensions~\cite{Zettlemoyer:2005,artzi2013weakly}.

Tests for any program will need to mention program-specific terms, which will need to be added by developers with no special linguistics background.
Experiments on \textsc{CCGBank} showed~\cite{hockenmaier2005ccgbank} that when training on most of lexicon, the unseen words in a held-out test set were primarily nouns (35.1\%) or transformations of nouns (e.g., adjectives, at 29.1\%). These are also the simplest categories for non-linguists to provide semantics for (types, objects, and predicates), suggesting that it should be possible to make the lexicon extendable for an individual program by normal developers without special linguistic background. (Similar experiments for a wide-coverage lexicon of German~\cite{hockenmaier2006creating} show over half of unknown words to be nouns, suggesting this feasibility extends beyond just English.)
\looseness=-1

A final key challenge --- common to any approach to relate formal and natural language --- is to find suitable evaluations for the efficacy of our technique.  Focusing on educational material we were only able to identify 26 explicit statements of an intended property in English, among 4 textbooks with explicit coverage of property-based testing~\cite{sttp,o2008real,hinojosa2013testing,lundin2015testing} (2 others taught \PBT without explicitly stating English for any test~\cite{wampler2014programming,chiusano2014functional}). Longer-term, the right evaluation approach is to ask developers currently using property-based tests to write natural language describing their test cases, or to seek access to existing closed-source pairings~\cite{arts2015testing}.

\bibliographystyle{ACM-Reference-Format}

\end{document}